\documentclass[pre,showpacs,preprint]{revtex4}

\usepackage{graphicx}
\usepackage{natbib}
\usepackage{amssymb}
\usepackage{amsmath}
\usepackage{verbatim}

\begin{document}

\title{Aging of rotational diffusion in colloidal gels and glasses}

\author{S. Jabbari-Farouji $^{1,2}$,G H.\ Wegdam $^{1}$ and Daniel Bonn$^{1,3}$  }

\affiliation{$^{1}$ LPTMS, CNRS and Universit$\acute{e}$ Paris-Sud, UMR8626, Bat. 100, 91405 Orsay, France}
\affiliation{$^{2}$ Van der Waals-Zeeman Institute, Institute of Physics (IoP) of the Faculty of Science (FNWI) University of
Amsterdam,  1098 XH Amsterdam, the Netherlands}
\affiliation{$^{3}$ Laboratoire de Physique Statistique de l'ENS, 24 rue Lhomond, 75231 Paris Cedex 05, France}
\date{\today}

\date{\today}

\begin{abstract}

We study the  rotational diffusion of aging  Laponite suspensions  for a wide range of concentrations using depolarized dynamic light scattering. The measured  orientational correlation functions undergo an ergodic to non-ergodic transition that is characterized by a concentration-dependent ergodicity-breaking time. We find that the relaxation times associated with rotational degree of freedom  as a function of waiting time, when scaled with their ergodicity-breaking time, collapse on two distinct master curves. These master curves are similar to those previously found for the translational dynamics; The two different classes of behavior were attributed to colloidal gels and glasses. Therefore, the aging dynamics of rotational degree of freedom  provides another signature of the distinct  dynamical behavior  of colloidal gels and glasses.
\\
{\it keywords: Rotational diffusion, Aging, colloidal glass, colloidal gel and  decoupling }
\end{abstract}

\maketitle

\section{Introduction}

Understanding the nature of arrested states of matter such as colloidal gels and glasses has been an active and challenging field of
research in recent decades \cite{Cates} and is still far from being well-understood. What makes these ubiquitous soft glassy materials a fascinating subject of study is their complex dynamical behavior that differs from that of equilibrium liquids.
What makes studying their dynamics interesting is the observation of universal dynamical features  in  different  systems, such as molecular, colloidal, metallic and ionic  glasses \cite{DH}. A very important common feature in such glassy  materials is a dramatic slowing down of  the diffusion of particles or molecules concomitant with a huge increase of viscosity on approaching the glass transition. The slowing down of dynamics expresses itself as a   non-exponential relaxation of the translational and  rotational degrees of freedom in supercooled liquids and non-decaying plateaus deep into the glassy phase. Another  significant feature is \emph{aging} i.e., the evolution of  the physical properties of the system with waiting time, the time elapsed since the quench into glassy phase.

An important question that arises in this respect is how the motion of  orientational and  translational degrees of freedom are influenced upon  the transition from a liquid-like to a  disordered solid-like state and if the slowing down of the two degrees of freedom are correlated. In a low-viscosity liquid far from the supercooled regime, the translational and rotational diffusion coefficients  of particles follow the Stokes-Einstein and Debye-Stokes-Einstein  equations, respectively, that relate the diffusion coefficients to temperature and  viscosity \cite{SE,DSE}. According to these relations the translational and rotational diffusion coefficients are not independent and  should both have a linear temperature dependence and be proportional to the inverse of the viscosity. While there have been some comprehensive studies on rotational motion of  several molecular glass formers \cite{decoupling1,decoupling2,decoupling3,decoupling4}, there are fewer studies on rotational dynamics of glassy colloidal systems \cite{Sara,Levitz2005,rotation-colloid,Sarasoft}.  Here, we  present   the results of  a systematic investigation of aging of the  translational and rotational diffusion of charged  colloidal platelets of Laponite  studied for a wide range of concentrations and as a function of waiting time.

 Laponite is a synthetic clay  \cite{Laponite}  that is widely used as a rheology-modifier in industrial  applications.  Laponite particles, when suspended in water,  form  a suspension  that evolves from a low-viscosity water-like liquid to  a highly viscoelastic fluid over time \cite{Mourchid,kroon,glass,Bonn1,aging,Bonn2,Sara-MR,Italian,PRL}. In other words, the suspension exhibits a strong aging behavior and undergoes an ergodic to non-ergodic transition. This behavior is observed at moderately low  volume fractions $\phi \approx 0.001-0.015$ \cite{Italian,PRL}.  The aging dynamics of translational diffusion and structure factor in both ergodic and non-ergodic regimes have been studied extensively over the past decade \cite{Italian,PRL,Italiannew,Lapreview} revealing  a dramatic slowing down of the translational motion. Interestingly, the intermediate scattering function measurements of Laponite suspensions have revealed that low and high concentration samples evolve in a distinctly different manner \cite{Italian,PRL,Italiannew}. Based on different aging behaviors of the translational diffusion as well as the structure factor,  the non-equilibrium states of Laponite suspensions in pure water have been identified to be a colloidal gel at low concentrations $C < 1.5 $ wt\% and a colloidal glass at higher concentrations $C > 1.5$ wt\% \cite{PRL,Italiannew,Lapreview}.
 
  Here, we study the  behavior of rotational diffusion when varying the concentration from the gel to the glass region. In prior works \cite{Sara,Sarasoft}, we have studied the aging dynamics of rotational diffusion in a colloidal glass of Laponite at concentrations around 3 wt \%. We showed that orientational correlation function of the system also shows a non-exponential decay and  the rotational relaxation time dramatically slows down during the aging; this happens at a rate faster than that of translational degree and viscosity. Here, we would like to know how the aging of rotational dynamics is correlated with the translational dynamics  during the aging as we vary the concentration and consequently if the rotational dynamics of colloidal gels and glasses can also be classified into two distinct groups (gels and glasses).  To this end,
  we perform a systematic study of the orientational correlations for a wide range of concentrations. We observe that the orientational dynamics of both colloidal gels and glasses slows down during the aging. However the rotational dynamics of colloidal gels and glasses behave distinctly differently, in line with observations for the translational dynamics, thus providing another criterion for discerning the two states. Also we find that in both colloidal gels and glasses, the orientational relaxation time grows at a faster rate than the translational one, suggesting that orientation and translation become decoupled upon approaching the non-ergodic state.

The remainder of paper is organized as follows. First, we present the experimental aspects of the sample preparation and dynamic light scattering measurements in section II. Subsequently, we briefly review, the theory of dynamic light scattering from anisotropic particles in section III.  Section IV  is devoted to the results of our measurements on rotational dynamics for colloidal gels and glasses of Laponite. In the last section V,  we summarize our main findings and discuss our results in relation to experiments on  other glass-forming materials.

\section{Experimental}

\subsection{Materials}
The Laponite that we used for our experiments is Laponite XLG that consists of charged platelets of  average
diameter of 30 nm and  thickness  of 1.2 nm \cite{Laponite}. Laponite can absorb water,
increasing its weight up to 20\%. Therefore, we first dried it in
an oven at $100^{o}$C for one week and subsequently stored it in a
desiccator.

We prepare a number of  samples with different concentrations of
Laponite in pure water.  Laponite dispersions are prepared in
ultrapure Millipore water $(18.2 M\Omega cm^{-1})$ and are
stirred vigorously by a magnetic stirrer for 1.5  h to make sure
that the Laponite particles are fully dispersed. The dispersions
are filtered using Millipore Millex AA $0.8\mu  m$ filter units to
obtain a reproducible initial state \cite{glass}. This instant
defines the zero of waiting time, $t_{w}=0$.

\subsection{Dynamic Light scattering setup}

Our  dynamic light scattering setup (ALV) is based on a He-Ne
laser ($\lambda = 632.8 nm$, 35 mW) and avalanche photodiodes as
detectors. An ALV-60X0 correlator directly computes the intensity
correlation functions
$g(q,t)=\frac{<I(q,t)I(q,0)>}{<I(q,0)>^{2}}$,  at a scattering wave
vector $q=\frac{4\pi n}{\lambda}\sin (\frac{\Theta}{2})$, in which
$\Theta$ is the scattering angle.

In our experiments the polarization of scattered light is detected
in two modes: The VV mode in which the polarization of the incident
and scattered light are both vertical  and the VH mode in
which the polarization of the scattered light is horizontal and
perpendicular to the vertical polarization of incoming light. The
VV and VH intensity correlation functions were measured at a fixed
scattering angle $\Theta=90^{o} $  ($q=1.87 \times 10^7$ m$^{-1}$) regularly at a rate depending
on the speed of aging of Laponite suspensions.

\section{Dynamic light scattering from anisotropic particles}

 The  anisotropy in shape of particles  gives rise to an anisotropic polarizability
tensor. For particles with axial symmetry  such as disks and rods,
 the eigenvectors  of polarizability tensor correspond to the directions perpendicular
and parallel to the symmetry
axis  and their resultant eigenvalues are  $\gamma_{\bot}$ and   $\gamma_{||}$, respectively  \cite{Pecora}. The total electric field scattered by such particles for linearly polarized incident light in the vertical direction has two components. The
first is the vertically \textit{polarized} component $E_{VV}$ with
an amplitude proportional to the average polarizability,
$\gamma=(\gamma_{||}+2\gamma_{\bot})/3$. The second one is the
horizontal \textit{depolarized} component $E_{VH}$. Its amplitude
is proportional to the intrinsic particle anisotropy
$\beta=\gamma_{||}-\gamma_{\bot}$, which is the difference between
the polarizabilities parallel and perpendicular to the optical
axis. In depolarized dynamic light scattering (DDLS), one measures
the correlation functions of the scattered light intensity   whose
polarization (horizontal) is perpendicular  to the polarization of
incident light (vertical), i.e. $g_{VH}$ , as opposed to the $g_{VV}$
for which the polarization of scattered and incident light
are both vertical. These intensity correlations are related to their corresponding intermediate scattering functions:
\begin{eqnarray} \label{eq:VH}
 f_{VV}(q,t)&=&\frac{\langle
 E_{VV}(\mathbf{q},t)E^*_{VV}(\mathbf{q},0)\rangle}{<I_{VV}(\mathbf{q})>} \\ \nonumber
 f_{VH}(q,t)&=&\frac{\langle
 E_{VH}(\mathbf{q},t)E^*_{VH}(\mathbf{q},0)\rangle}{<I_{VH}(\mathbf{q})>} \\ \nonumber
 \end{eqnarray}
through  the Siegert relation, i.e. $g_{VV (H)}=1+ C |f_{VV(H)}|^2$, where $C$ is a set-up dependent coherence factor \cite{Pecora}.
Assuming that the suspensions  are dilute enough so that
orientations and positions of different particles are
uncorrelated, one can obtain the VV and VH intermediate scattering functions in terms of the translational and rotational diffusion coefficients and average polarizability   $\gamma$ and anisotropy  of polarizability $\beta$  \cite{Pecora}.  The resulting
intermediate scattering functions factorize into a product of a
 purely local (not $q$-dependent) orientational correlation and  a $q$-dependent
translational correlation and can be written as \cite{Pecora}:
 \begin{eqnarray}\label{eq:VVH}
 f_{VV}(q,t)=\frac{[\gamma^{2}+\frac{4}{45}\beta^{2}\exp(-6D_rt)]F_{s}(q,t)}{\gamma^{2}+\frac{4}{45}\beta^{2}}
\\
f_{VH}(q,t)=F_{s}(q ,t)\exp(-6D_rt)
\end{eqnarray}
where
$F_{s}(q,t)=<\exp(i\mathbf{q}.[\mathbf{r_{i}}(0)-\mathbf{r_{i}}(t)])>$ is the  self  translational correlation function.  The contribution
of the rotational motion to the $VV$ correlation  is proportional
to $\frac{4}{45}\beta^2/ \gamma^{2} \propto I_{VH}/ I_{VV}$,  that turned out to be smaller than 0.02 in our experiments.
Therefore we can neglect the rotational contribution to $f_{VV}$ in our data. Hence the dynamics of $f_{VV}$ mainly reflects  the
translational diffusion of the particles, while $f_{VH}$ is
determined by both translational and rotational motion.

The translational and rotational diffusion coefficients of dilute
solutions of hard disks can be obtained as the limiting case of the general formula
for diffusion of oblate spheroids with major semi-axis $b$ and minor
semi-axis $a$ \cite{Pecora}. For disks taking $a=0$ and $b=R$, we  can obtain in the infinite-dilution limit, the translational diffusion coefficients parallel  and perpendicular to the symmetry axis of the disk,  respectively, as: $D_{||}^t=\frac{k_BT}{16 \eta R}$ and $D_{\bot}^t=\frac{3k_BT}{32 \eta R}$ giving an average translational diffusion $D_{0}^t=1/3(D_{|| }+2D_{\bot })=\frac{kT}{12\eta R}$; the rotational diffusion coefficient follows as  $D_{0}^r=\frac{3kT}{32\eta R^{3}}$ .  In DLS
experiments in the VV mode, we measure the  average translational
diffusion $D^t$ that for dilute suspensions should be $D_{0}^t$.
From depolarized DLS (VH mode),  we can  extract the  rotational diffusion coefficient $D^r$.

\section{Aging dynamics of rotation and translation   in gels and  glasses of Laponite suspensions}
\begin{figure}[ht]
\includegraphics [scale=0.52]{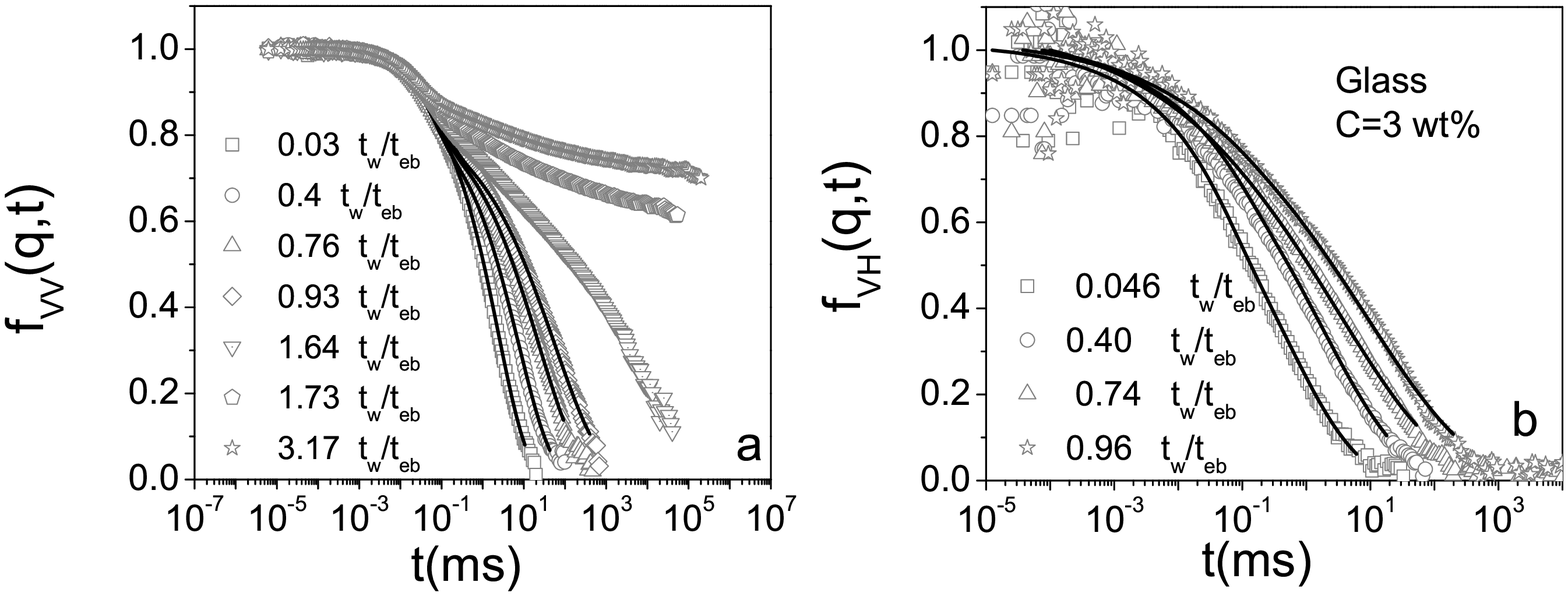}
\includegraphics [scale=0.52] {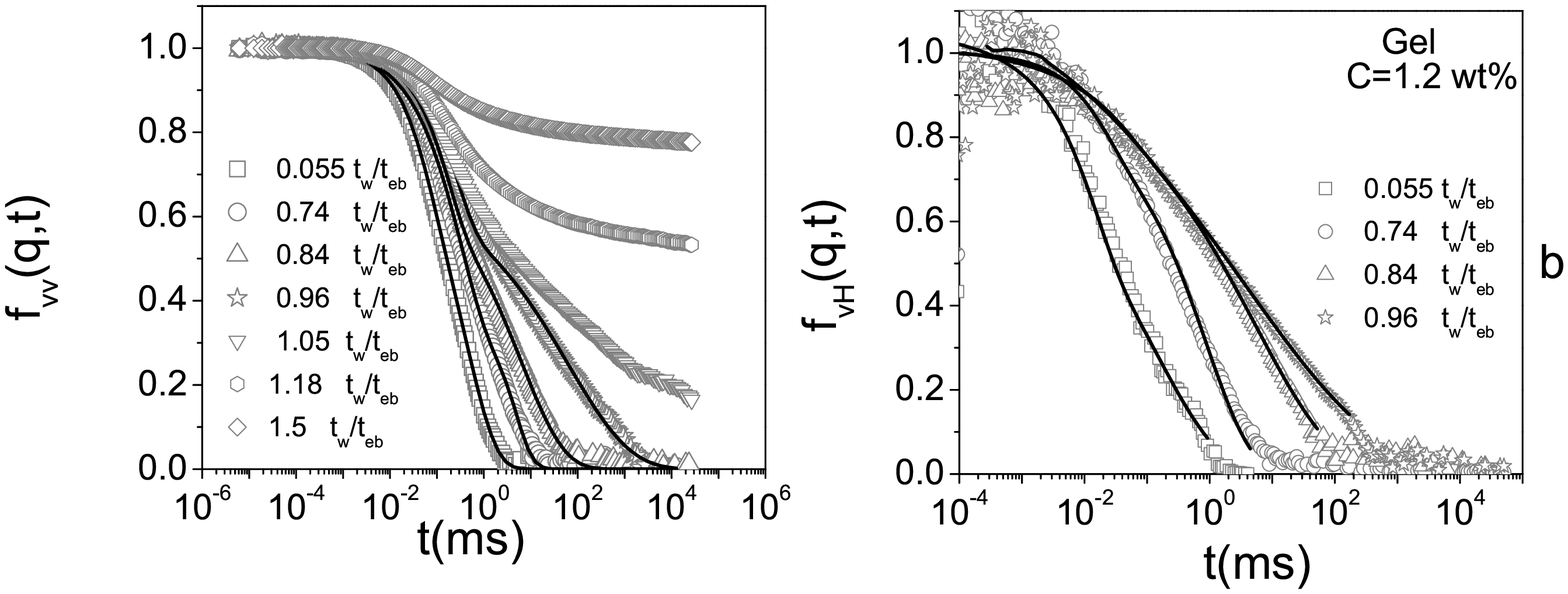}
\caption{
 Evolution of polarized (VV) and depolarized (VH) intensity correlation functions (symbols) and
their corresponding fits (solid lines) to Eq. (\ref{eq:stretch1})
a) for a glass (Laponite 3 wt\%, pure water) and b) for a gel
(Laponite 1.2 wt\%, pure water) measured at a scattering angle of
$90^{o}$. The waiting times are shown in the legends. Note that
here only the correlations in the ergodic regime of aging are
fitted. The ergodicity-breaking times are: $t_{eb}=450$ min for
Laponite 3wt\% and $t_{eb}=18$ days for Laponite 1.2wt\%.
}\label{fig1}
\end{figure}

Measuring the intensity correlations of scattered light from
 aging Laponite dispersions, one always observes two
regimes of aging in the evolution of the intensity correlations  \cite{kroon,glass,PRL}. In the first regime of aging, the translational correlation functions decay to zero within the experimental accessible time scales. In  the second regime of aging,  the ensemble-averaged correlations do not decay to zero and a plateau in intermediate scattering function develops (see Fig. \ref{fig1}a). The waiting time for which the correlation the  functions no longer decay to zero within the experimental time-scale or, equivalently, the moment when
the time-averaged correlation functions are no longer equal to their ensemble-averaged values,
defines the ergodicity-breaking time $t_{eb}$. Studying the aging dynamics of VV and VH correlation functions simultaneously, we find  that the VH correlations become non-ergodic at nearly the same time as VV correlations. Therefore,  we take the same $t_{eb}$ value for both VV and VH correlations. To be able to quantify the relaxation times for rotational and translational diffusion, here we only focus on aging dynamics in the first, ergodic, regime of aging.

\begin{figure}[t]
\includegraphics [scale=0.60]{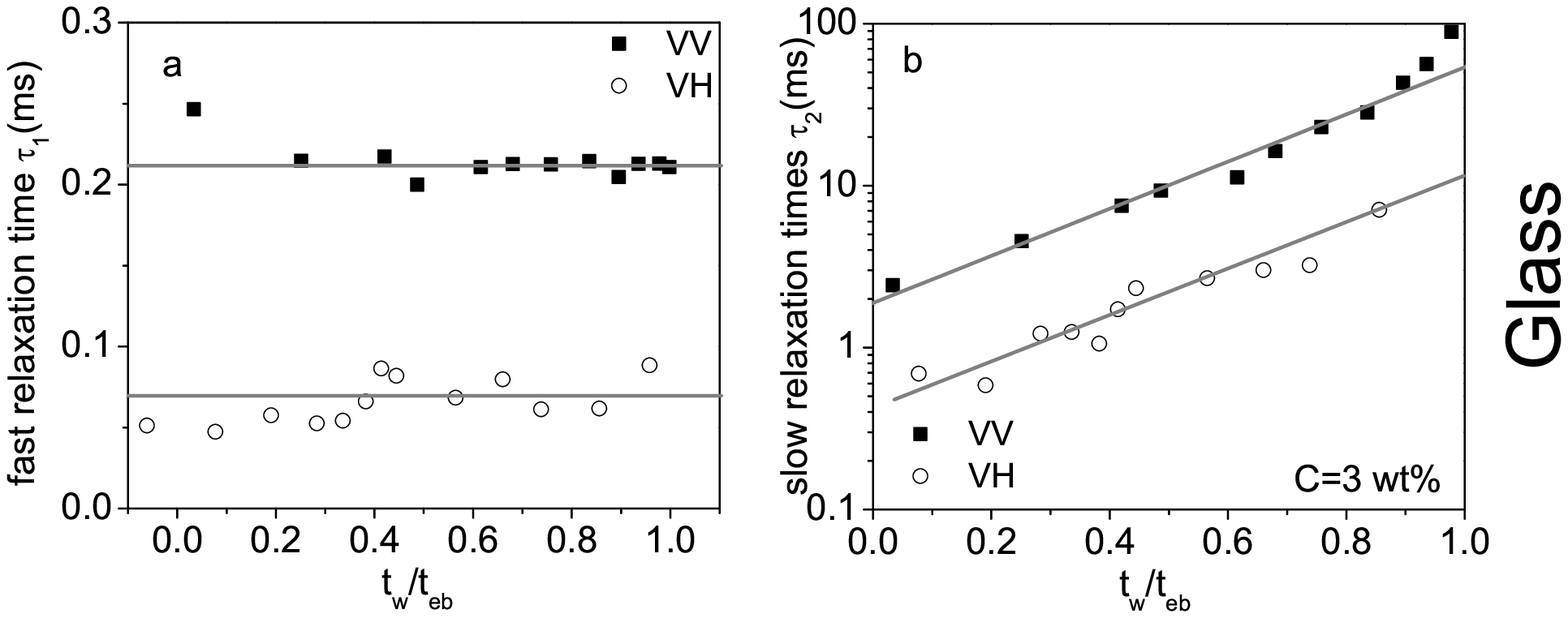}
\includegraphics [scale=0.60] {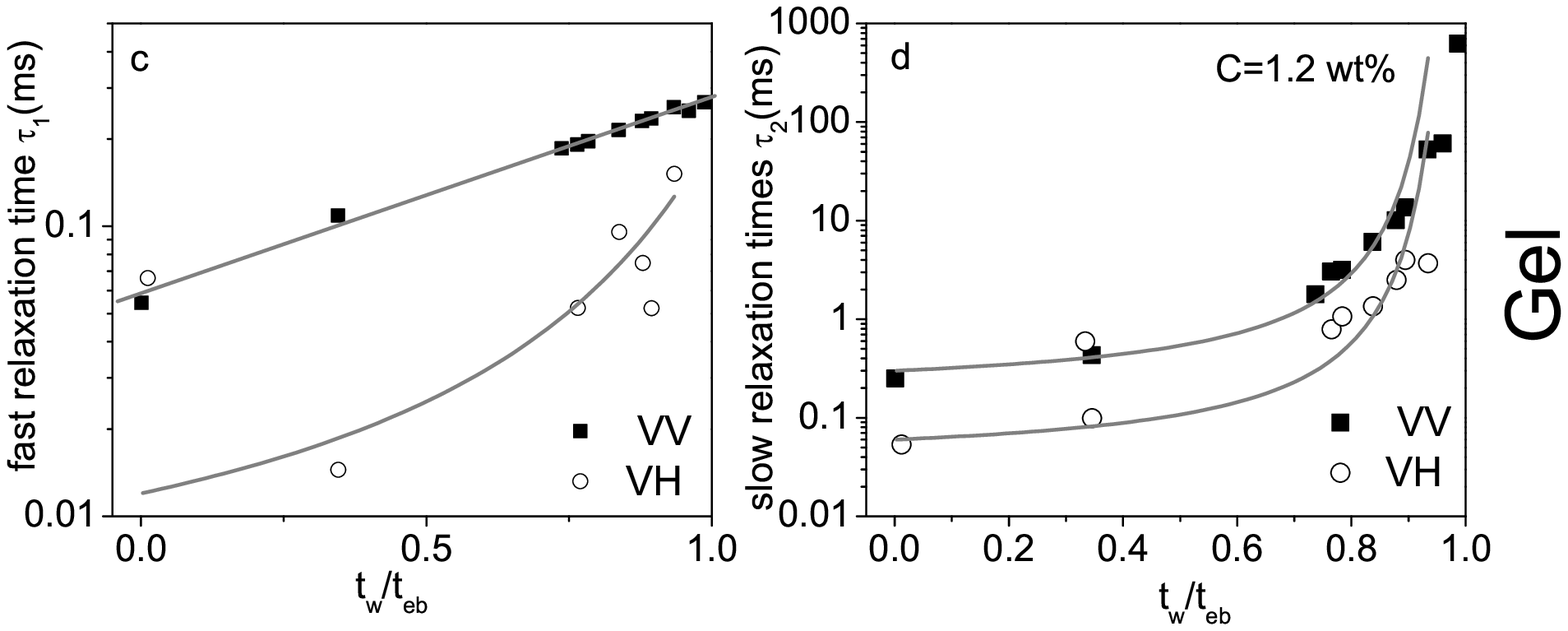}
\includegraphics [scale=0.60]{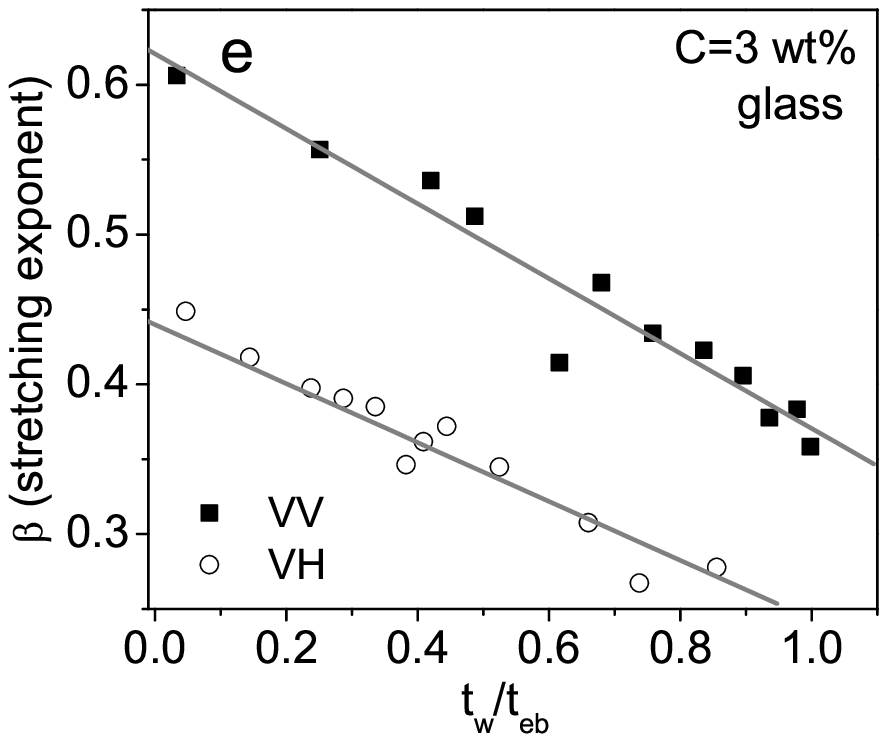}
\includegraphics [scale=0.20] {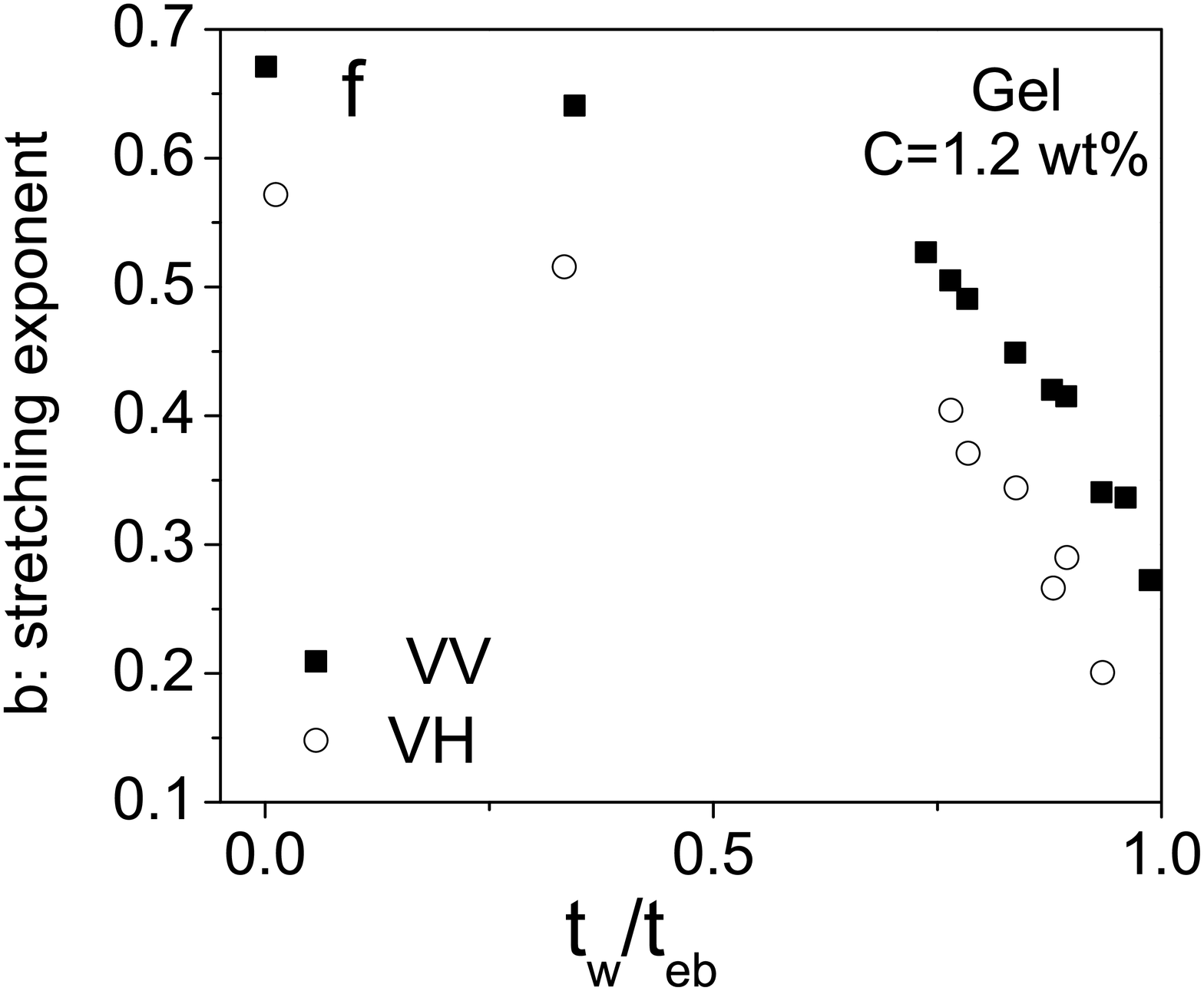}
\caption{ Evolution of fast ($\tau_1$) and slow ($\tau_2$)  relaxation
times and the stretching exponents $\beta$ of VV and VH
correlation functions plotted versus scaled waiting time
$t_w/t_{eb}$ for a
 glass (Laponite 3 wt\%, pure water) and a gel
(Laponite 1.2 wt\%, pure water). The lines show the fits of
relaxation times with the general form $\tau_{i}(t_w)= \tau_0
\exp(B \frac{t_w}{t_w^{\infty}-t_w})$ \cite{Italian}. }\label{fig2}
\end{figure}

Figure \ref{fig1} shows the evolution of VV and VH intermediate
scattering functions for a sample of low   and a sample of
high concentration  measured at different stages of aging. As can be observed in the figure, for both  the VV and VH  intermediate scattering functions a two-step relaxation  can be observed \cite{Bonn1,Italian,PRL} and in both samples, the average relaxation times of VV and VH correlations increase with increasing waiting time.  In order to consistently describe these two step  relaxations quantitatively, we
fit the normalized correlation functions, both VV and VH, by the
sum of an exponential and a stretched exponential  \cite{Bonn1,Italian,Sara}:
\begin{equation}\label{eq:stretch1}
 f(q,t)-1=A \exp(-t/\tau_{1})+(1-A)\exp(-(t/\tau_{2})^{\beta}).
   \end{equation}
where $A$ determines the relative contribution of exponentially decaying fast relaxation time $\tau_1$ and $b$ is known  as the stretching exponent characterizing the broadness of the distribution of slow relaxation modes  that contribute to the correlation function. The smaller $b$ is, the broader  the distribution of the slow relaxation times is. The slow relaxation time $\tau_2$ gives us an idea about the mean relaxation time that can be obtained as $\tau_m= \tau_2/b \Gamma (1/b)$, where $\Gamma$ is the gamma function. To ensure the accuracy of the extracted value for
   short-time diffusion and constrain the fitting procedure,
   we first determined $\tau_{1}$ independently, using a
   linear fit of $\ln(f(q,t)-1)$ for short times
  ( $t<0.005$ ms), corresponding to the short-time diffusion. We  subsequently use the  extracted value of $\tau_1$ to obtain $A$, $b$ and $\tau_2$ from fitting with in Eq. (\ref{eq:stretch1}).

\begin{figure}[t]
\begin{center}
\includegraphics [scale=0.60]{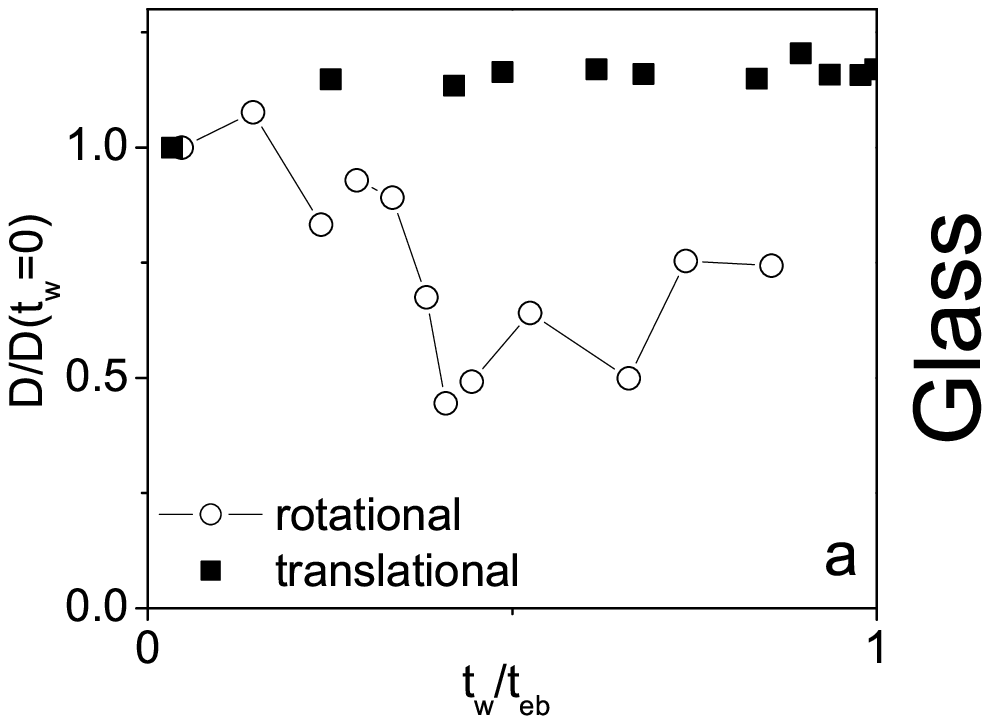}
\includegraphics [scale=0.60]{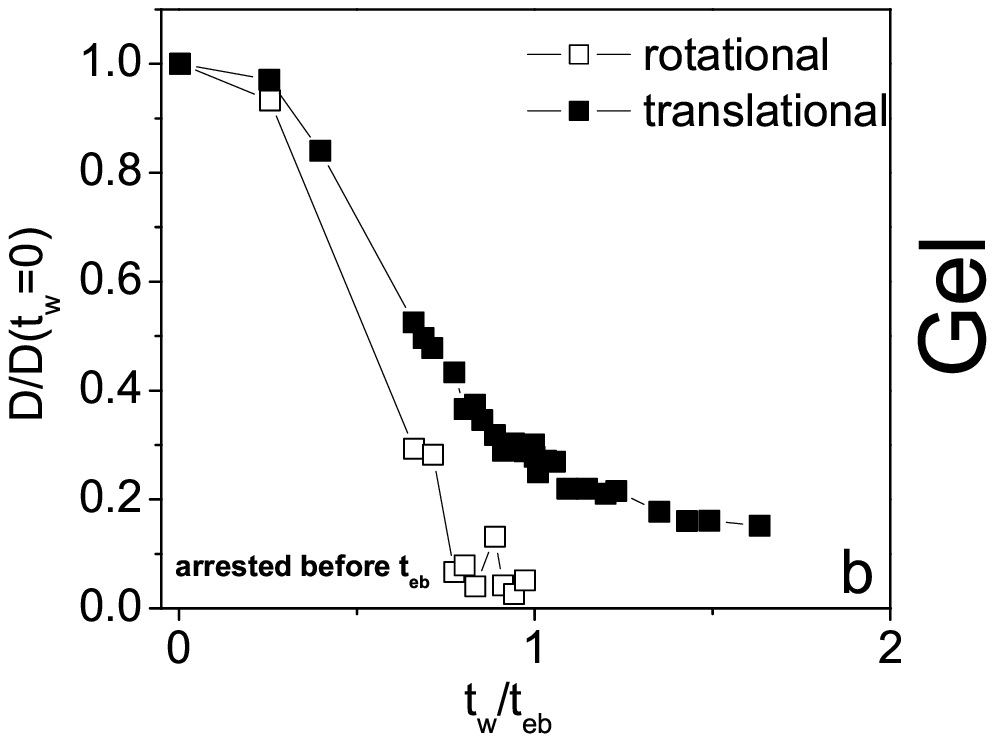}
\caption{Short-time  translational and rotational diffusion
coefficients normalized to their values at $t_w\approx0 $ as a
function of scaled aging time $t_w/t_{eb}$ a) in a glass of
Laponite C=3.2 wt\% b) in a gel of Laponite  C= 1.2
wt\%}\label{fig3}
\end{center}
\end{figure}

The results of fits of the VV and VH correlations for both  samples are depicted in Fig. \ref{fig2}.
For the high-concentration  sample  the fast relaxation times  are roughly
independent of the waiting time (Fig. \ref{fig2} a), while  they  grow  in the low-concentration sample, as presented in Fig.\ \ref{fig2} c.  The short-time translational $D_s^t$ and rotational $D_s^r$ diffusion coefficients can be extracted from the fast relaxation times as $D_s^t=1/ ( \tau_1^{VV}q^2)$ and $D_s^r= 1/(6 \tau_1^{VH})-1/( 6 \tau_1^{VV}q^2)$. The resulting short-time diffusion coefficients normalized to their initial values ($t_w \approx 0$) for both   samples  are  plotted in Fig.\ \ref{fig3}a and b respectively.  Fig.\ \ref{fig3} b clearly demonstrates
that the short-time rotational diffusion coefficient $D_s^r$ in the low-concentration  decreases at a faster
rate than the translational  one $D_s^t$. In contrast, for the high-concentration sample both short-time  translational and rotational diffusion coefficients show only a weak dependence on the waiting time. The short-time rotational diffusion already indicates that the aging of two types of samples are different.

 We  now turn to the behavior of the slow relaxation times.  Fig. \ref{fig2} (b) is similar to what was reported in previous works for the colloidal glass of Laponite \cite{Sara}: the slow relaxation times associated with translational and rotational degrees of freedom grow exponentially  with waiting time  and remain parallel to each other.  In the low concentration samples that were previously identified as gels \cite{PRL,Italiannew,Lapreview} the slow relaxation time  grows faster than exponentially  for both VV and VH  correlations as shown in  Fig.\ \ref{fig2}(d). Looking at the stretching exponent $b$, we observe that $b$ decreases with waiting time for both VV and VH
correlations in both  the gel and the glass (Fig. \ref{fig2}(e) and (f)). However, the stretching exponents of the translational and rotational degree in the gel decreases at a faster rate  than in the glass, pointing to a different character of aging in  gels and glasses.

\begin{figure}[t]
\begin{center}
\includegraphics [scale=0.25]{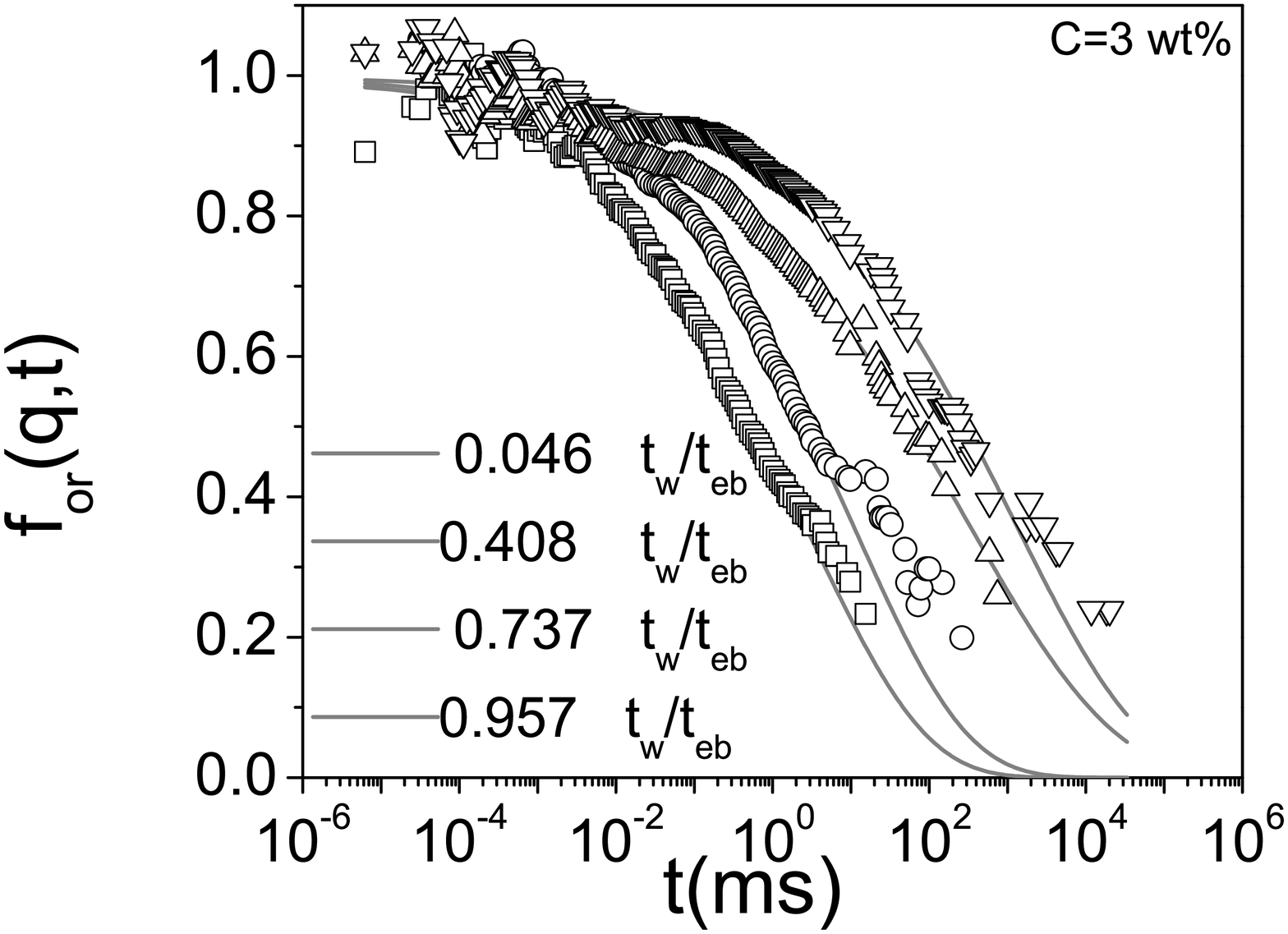}
\includegraphics [scale=0.25] {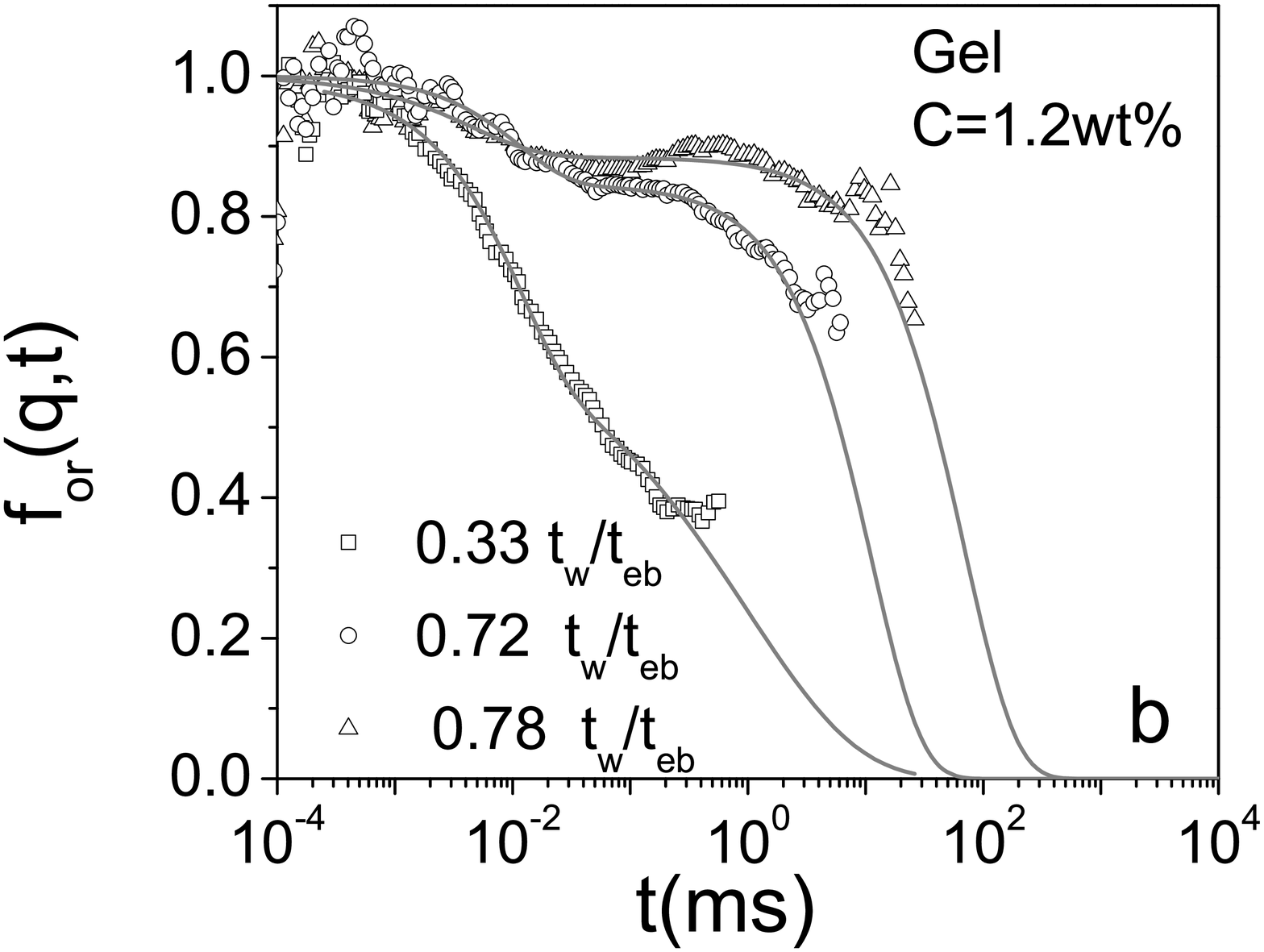}
\caption{
 The orientational correlation functions defined as $f_{VH}/f_{VV}$
 at different waiting times in
a glass (Laponite 3 wt\%, pure water) and a gel (Laponite 1.2
wt\%, pure water). The lines show the fits with the sum of single
and stretched exponential as in Eq. (\ref{eq:stretch1}).
  }\label{fig:orientation}
 \end{center}
\end{figure}

Before going further with  interpreting  these results, one should bear in mind that both translational and rotational degrees of
freedom contribute to the VH correlations in contrast to VV correlations that reflect almost exclusively the dynamics of translational motion. In order to gain a more direct insight into the rotational dynamics, we extract the orientational
correlation functions defined as the ratio $f_{or}(t)=\frac{f_{VH}(t)}{f_{VV}(t)}$. The obtained orientational correlations  for the glass
and the gel  at different stages of aging are depicted in Fig.\ \ref{fig:orientation}.  The orientational correlations of both gel
and glass   can  also be fitted with  Eq.\ (\ref{eq:stretch1}).

\begin{figure}[t]
\begin{center}
\includegraphics [scale=0.26]{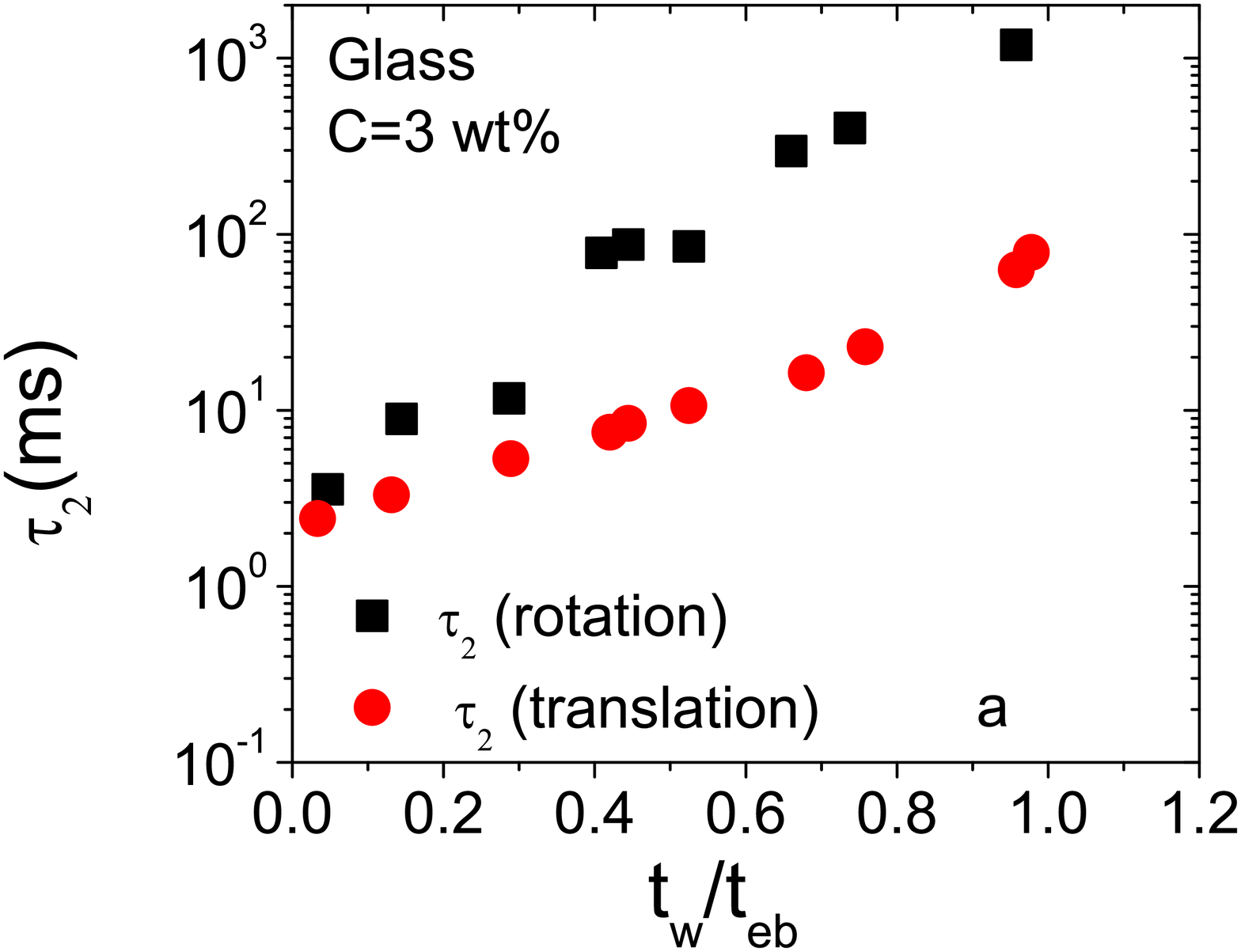}
\includegraphics [scale=0.25]{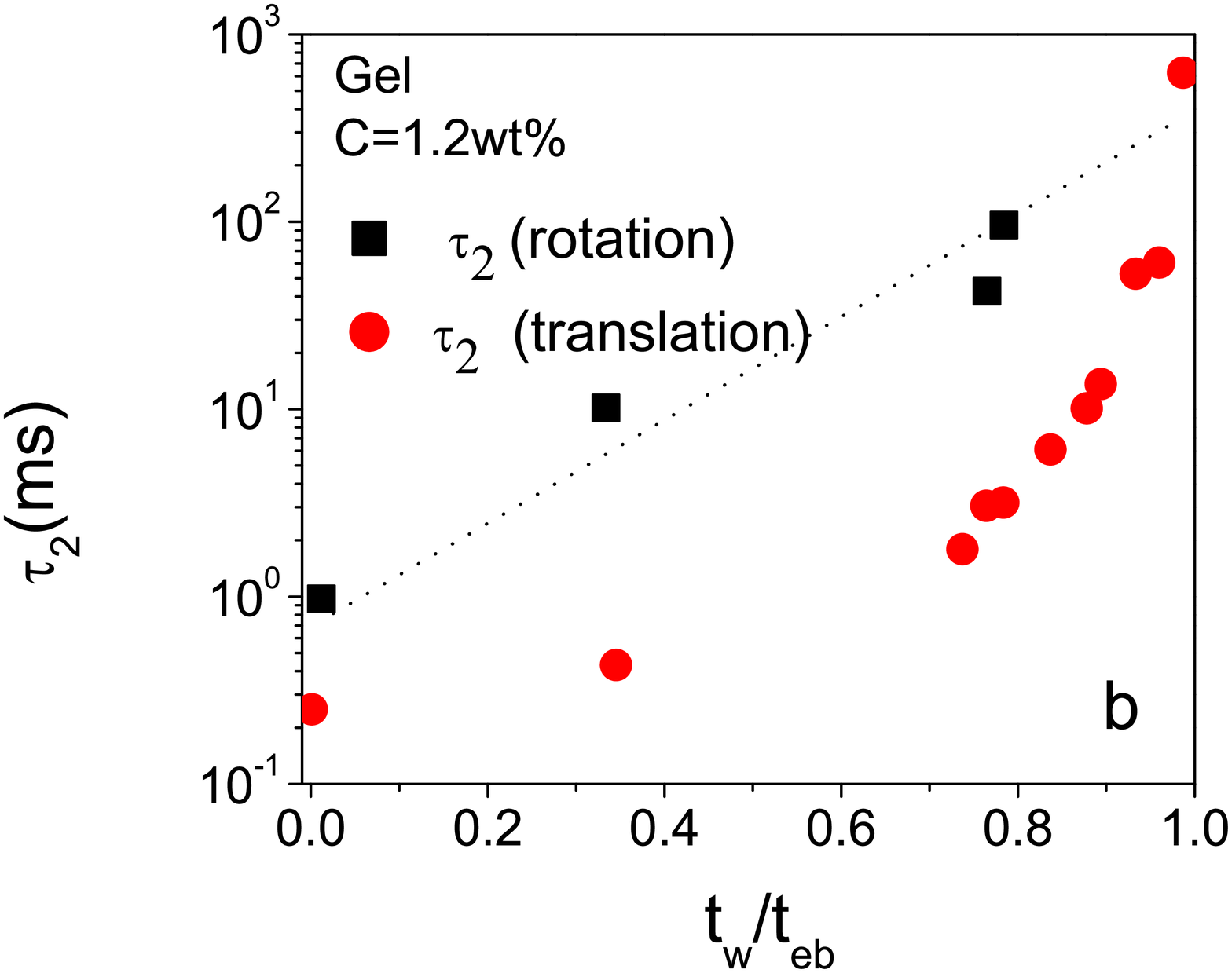}
\caption{ The orientational and translational relaxation times versus scaled waiting time
 measured for a Laponite suspension of a) $C= 3$ wt\% (glass).  b)  $C=1.2$ wt\% (gel).
   }\label{fig:decoupling}
 \end{center}
\end{figure}

The slow relaxation times extracted from the orientational correlations are
presented in  Fig.\ \ref{fig:decoupling}, together with the relaxation times for the translational degree of freedom.  From this figure, we  observe that for both gel and glass samples the slow relaxation time of the orientational degree of freedom grows more rapidly with $t_w$  than the  translational one.  Furthermore, the orientational relaxation time of the gel ages in a manner that is different than in  the glass. The difference in the  slow relaxation times for orientational and  translational correlations  is an indication of the decoupling of translational degree of freedom from the translational degree of freedom. The fact that also the  orientational  correlation functions are also well described by a stretched exponential (Fig. \ref{fig:orientation}) indicates that the rotational degrees also have a broad distribution of relaxation times. Furthermore, the decrease of the stretching exponent with increasing waiting time expresses the fact that the distribution
of these relaxation times becomes wider as the system ages, again similar to earlier observations for the translational dynamics
\cite{Bonn1,Italian}.

\begin{figure}[h]
\includegraphics [scale=0.650]{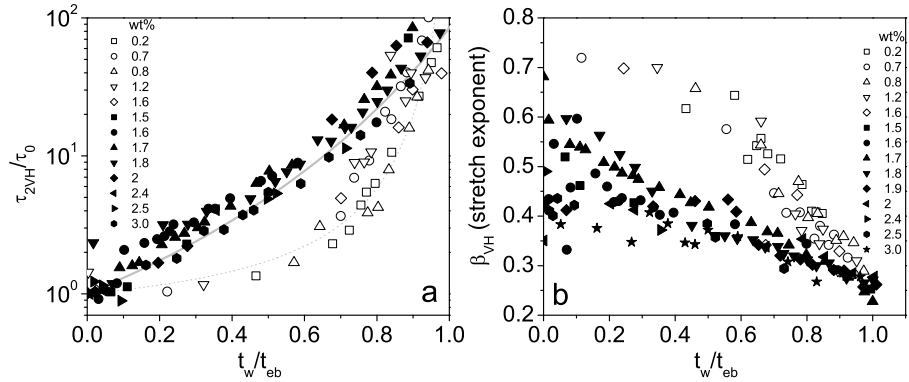}
\caption{The evolution of a) the slow relaxation time normalized
to its initial value $\tau_2/\tau_0$ b) the stretching exponent
$\beta$ of VH correlations versus scaled waiting time $t_w/t_{eb}$
for different Laponite samples. The colloid concentrations are
shown in the legend. Similarly to the translational degree of
freedom, the samples can be divided into two groups according to
the evolution of the slow relaxation times of VH correlations. In
both figures the open symbols correspond to gels, the filled
symbols to the glass. In panel a) the solid and dashed line
correspond to the glass and gel lines obtained from growth of
relaxation time of translational diffusion. }\label{fig:t2rot}
\end{figure}

Having investigated in detail the aging of translational and rotational diffusion in a gel and a glassy sample and characterizing their differences,  we now present our results for a wider range of samples studied.
We measured the VV and VH time correlation functions for a series of Laponite samples ranging from 0.2 wt \% to 3 wt \%. From these measurements, we find that  the aging behavior of rotational degree of freedom
between the low  (gel-like) and  high  (glassy) concentration samples is distinctly different. Plotting the quantities $\tau_2$ and $\beta$ for VH
correlations as a function of scaled waiting time $t_w/t_{eb}$ ,
we find that, similar to the slow relaxation time of VV
correlations, the data from all the samples collapse onto two distinct
master curves (Fig.\ \ref{fig:t2rot}). These two master curves can
be identified as gels and glasses in agreement with the
classification based on the aging of the translational degree of
freedom in \cite{PRL,Italian}.

\begin{figure}[t]
\includegraphics [scale=0.30]{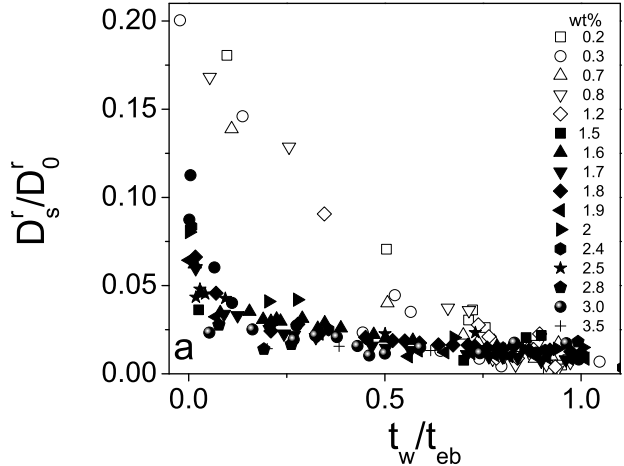}
\caption{a) The short time rotational diffusion coefficients
normalized to those of infinitely diluted hard disks of radius
$R=15$ nm: $D_{0}^r$.
}\label{fig7}
\end{figure}

Looking at the aging behavior of the short-time rotational diffusion of
all the samples, again we distinguish two groups of samples. Figure \ \ref{fig7} presents the short-time rotational diffusion
coefficients $D_{s}^{r}$, normalized to the rotational
diffusion of hard disks in the infinite dilution limit
$D_{0}^r=k_BT/32\eta R^3$, (taking $R=15 $nm: $D_{0}^r=1.1 \times 10^{5}$ s$^{-1}$) as a function of the scaled waiting time $t_w/t_{eb}$. We find  a moderate decrease of $D_{s}^{r}$ for the glassy
samples, while the short time rotational diffusion decreases  very significantly in the gels.

 As can be observed in Fig. \ref{fig7}, the rotational diffusion at the early stages of aging is considerably slower in glassy samples  and  is almost concentration-independent on approaching  the non-ergodic state.
 Therefore, one can deduce that  in the non-ergodic state the rotational diffusion of  both
gels and glasses is hindered. The rotational motion of particles in the glass is constrained from the very
beginning due to crowding of particles at high concentrations (glass), while in the gel the rotational motion of particles is obstructed as they become part of a network or clusters of particles.

\section{Conclusion}

We have studied the evolution of both the translational and
rotational dynamics during aging for a wide range of concentrations
of Laponite particles including both colloidal gels and glasses. We find that the the dynamics are qualitatively similar between
the two degrees of freedom in all the samples. Concomitant with
the slowing down of the translational motion, the rotational motion
slows down as well, however at a faster rate. More precisely,  both the short-time
rotational and translational diffusion decrease significantly
during  gel formation while these quantities remain almost constant during the formation of the glass.
The slow relaxation times, characterizing  the translational and orientational
degree of freedom at longer delay times, decrease at a faster rate in  the gel compared to the glass.
All these findings point to the different mechanisms involved in  gel and glass formation.
The  enormous slowing down of  short-time rotational  diffusion in the gel indicates that the rotational motion of particles in a gel  becomes very restricted as a result of the formation  of small clusters or network-like structures. Similarly to the translational degrees of freedom that was reported previously \cite{Italian, PRL,Lapreview}, the aging dynamics  of rotation in colloidal gels and glasses is different. One can observe two distinct routes of evolution for the rotational dynamics, providing us with another criterion  \cite{PRL,Italiannew,PREnew}to distinguish   colloidal gels (low concentrations) and glasses (high concentrations).

The  faster rate of slowing down of rotational motion compared to that  of translation in both colloidal gels and
glasses, points to a decoupling of translational and rotational motion during the aging of colloidal gels and glasses and is usually interpreted in terms of  dynamical heterogeneity \cite{decoupDynHet,Sarasoft}. The non-exponential behavior of both translational and
orientational correlations and their description by a stretched
rather than a simple exponential also  points to a broad distribution of
relaxation times, as thus also heterogeneous dynamics.  Our data also suggest that the decoupling is stronger in
spatially heterogeneous samples (gels) than the spatially homogeneous ones (glasses),
as the the difference in the aging behavior of rotational and
translational degrees of freedom is more pronounced in gels. These observations  agree with the existing literature where it is found that the dynamic heterogeneity plays a central role in the decoupling between rotational and translational diffusion \cite{decoupDynHet,decoupling1,decoupling2,decoupling3}.\\
\\

 \textbf{Acknowledgments} The research has been supported
by the Foundation for Fundamental Research on Matter (FOM), which
is financially supported by Netherlands Organization for
Scientific Research (NWO).  S. J-F. acknowledges further support of the foundation of "Triangle de la Physique" and IEF Marie-Curie fellowship.  LPS de l'ENS is UMR8550 of the CNRS, associated with the universities Paris 6 and 7.

\end {document}